\catcode`\@=11					



\font\fiverm=cmr5				
\font\fivemi=cmmi5				
\font\fivesy=cmsy5				
\font\fivebf=cmbx5				

\skewchar\fivemi='177
\skewchar\fivesy='60


\font\sixrm=cmr6				
\font\sixi=cmmi6				
\font\sixsy=cmsy6				
\font\sixbf=cmbx6				

\skewchar\sixi='177
\skewchar\sixsy='60


\font\sevenrm=cmr7				
\font\seveni=cmmi7				
\font\sevensy=cmsy7				
\font\sevenit=cmti7				
\font\sevenbf=cmbx7				

\skewchar\seveni='177
\skewchar\sevensy='60


\font\eightrm=cmr8				
\font\eighti=cmmi8				
\font\eightsy=cmsy8				
\font\eightit=cmti8				
\font\eightbf=cmbx8				

\skewchar\eighti='177
\skewchar\eightsy='60


\font\ninei=cmmi9
\font\ninesy=cmsy9

\skewchar\ninei='177
\skewchar\ninesy='60


\font\tenrm=cmr10				
\font\teni=cmmi10				
\font\tensy=cmsy10				
\font\tenex=cmex10				
\font\tenit=cmti10				
\font\tensl=cmsl10				
\font\tenbf=cmbx10				
\font\tentt=cmtt10				
\font\tenss=cmss10				
\font\tensc=cmcsc10				
\font\tenbi=cmmib10				

\skewchar\teni='177
\skewchar\tenbi='177
\skewchar\tensy='60

\def\tenpoint{\ifmmode\err@badsizechange\else
	\textfont0=\tenrm \scriptfont0=\sevenrm \scriptscriptfont0=\fiverm
	\textfont1=\teni  \scriptfont1=\seveni  \scriptscriptfont1=\fivemi
	\textfont2=\tensy \scriptfont2=\sevensy \scriptscriptfont2=\fivesy
	\textfont3=\tenex \scriptfont3=\tenex   \scriptscriptfont3=\tenex
	\textfont4=\tenit \scriptfont4=\sevenit \scriptscriptfont4=\sevenit
	\textfont5=\tensl
	\textfont6=\tenbf \scriptfont6=\sevenbf \scriptscriptfont6=\fivebf
	\textfont7=\tentt
	\textfont8=\tenbi \scriptfont8=\seveni  \scriptscriptfont8=\fivemi
	\def\rm{\tenrm\fam=0 }%
	\def\it{\tenit\fam=4 }%
	\def\sl{\tensl\fam=5 }%
	\def\bf{\tenbf\fam=6 }%
	\def\tt{\tentt\fam=7 }%
	\def\ss{\tenss}%
	\def\sc{\tensc}%
	\def\bmit{\fam=8 }%
	\rm\setparameters\setbaselines\fi}


\font\twelverm=cmr12				
\font\twelvei=cmmi12				
\font\twelvesy=cmsy10	scaled\magstep1		
\font\twelveex=cmex10	scaled\magstep1		
\font\twelveit=cmti12				
\font\twelvesl=cmsl12				
\font\twelvebf=cmbx12				
\font\twelvett=cmtt12				
\font\twelvess=cmss12				
\font\twelvesc=cmcsc10	scaled\magstep1		
\font\twelvebi=cmmib10	scaled\magstep1		

\skewchar\twelvei='177
\skewchar\twelvebi='177
\skewchar\twelvesy='60

\def\twelvepoint{\ifmmode\err@badsizechange\else
	\textfont0=\twelverm \scriptfont0=\eightrm \scriptscriptfont0=\sixrm
	\textfont1=\twelvei  \scriptfont1=\eighti  \scriptscriptfont1=\sixi
	\textfont2=\twelvesy \scriptfont2=\eightsy \scriptscriptfont2=\sixsy
	\textfont3=\twelveex \scriptfont3=\tenex   \scriptscriptfont3=\tenex
	\textfont4=\twelveit \scriptfont4=\eightit \scriptscriptfont4=\sevenit
	\textfont5=\twelvesl
	\textfont6=\twelvebf \scriptfont6=\eightbf \scriptscriptfont6=\sixbf
	\textfont7=\twelvett
	\textfont8=\twelvebi \scriptfont8=\eighti  \scriptscriptfont8=\sixi
	\def\rm{\twelverm\fam=0 }%
	\def\it{\twelveit\fam=4 }%
	\def\sl{\twelvesl\fam=5 }%
	\def\bf{\twelvebf\fam=6 }%
	\def\tt{\twelvett\fam=7 }%
	\def\ss{\twelvess}%
	\def\sc{\twelvesc}%
	\def\bmit{\fam=8 }%
	\rm\setparameters\setbaselines\fi}


\font\fourteenrm=cmr12	scaled\magstep1		
\font\fourteeni=cmmi12	scaled\magstep1		
\font\fourteensy=cmsy10	scaled\magstep2		
\font\fourteenex=cmex10	scaled\magstep2		
\font\fourteenit=cmti12	scaled\magstep1		
\font\fourteensl=cmsl12	scaled\magstep1		
\font\fourteenbf=cmbx12	scaled\magstep1		
\font\fourteentt=cmtt12	scaled\magstep1		
\font\fourteenss=cmss12	scaled\magstep1		
\font\fourteensc=cmcsc10 scaled\magstep2	
\font\fourteenbi=cmmib10 scaled\magstep2	

\skewchar\fourteeni='177
\skewchar\fourteenbi='177
\skewchar\fourteensy='60

\def\fourteenpoint{\ifmmode\err@badsizechange\else
	\textfont0=\fourteenrm \scriptfont0=\tenrm \scriptscriptfont0=\sevenrm
	\textfont1=\fourteeni  \scriptfont1=\teni  \scriptscriptfont1=\seveni
	\textfont2=\fourteensy \scriptfont2=\tensy \scriptscriptfont2=\sevensy
	\textfont3=\fourteenex \scriptfont3=\tenex \scriptscriptfont3=\tenex
	\textfont4=\fourteenit \scriptfont4=\tenit \scriptscriptfont4=\sevenit
	\textfont5=\fourteensl
	\textfont6=\fourteenbf \scriptfont6=\tenbf \scriptscriptfont6=\sevenbf
	\textfont7=\fourteentt
	\textfont8=\fourteenbi \scriptfont8=\tenbi \scriptscriptfont8=\seveni
	\def\rm{\fourteenrm\fam=0 }%
	\def\it{\fourteenit\fam=4 }%
	\def\sl{\fourteensl\fam=5 }%
	\def\bf{\fourteenbf\fam=6 }%
	\def\tt{\fourteentt\fam=7}%
	\def\ss{\fourteenss}%
	\def\sc{\fourteensc}%
	\def\bmit{\fam=8 }%
	\rm\setparameters\setbaselines\fi}


\font\seventeenrm=cmr10 scaled\magstep3		


\newdimen\rp@
\newcount\@basestretchnum
\newskip\@baseskip
\newskip\headskip
\newskip\footskip


\def\setparameters{\rp@=.1em
	\headskip=24\rp@
	\footskip=\headskip
	\delimitershortfall=5\rp@
	\nulldelimiterspace=1.2\rp@
	\scriptspace=0.5\rp@
	\abovedisplayskip=10\rp@ plus3\rp@ minus5\rp@
	\belowdisplayskip=10\rp@ plus3\rp@ minus5\rp@
	\abovedisplayshortskip=5\rp@ plus2\rp@ minus4\rp@
	\belowdisplayshortskip=10\rp@ plus3\rp@ minus5\rp@
	\normallineskip=\rp@
	\lineskip=\normallineskip
	\normallineskiplimit=0pt
	\lineskiplimit=\normallineskiplimit
	\jot=3\rp@
	\setbox0=\hbox{\the\textfont3 B}\p@renwd=\wd0
	\skip\footins=12\rp@ plus3\rp@ minus3\rp@
	\skip\topins=0pt plus0pt minus0pt}


\def\setbaselines{\maxdepth=4\rp@\baselinestretch=\@basestretchnum}


\def\baselinestretch{\afterassignment\@basestretch\@basestretchnum}
\def\@basestretch{%
	\@baseskip=12\rp@ \divide\@baseskip by1000
	\normalbaselineskip=\@basestretchnum\@baseskip
	\baselineskip=\normalbaselineskip
	\bigskipamount=\the\baselineskip
		plus.25\baselineskip minus.25\baselineskip
	\medskipamount=.5\baselineskip
		plus.125\baselineskip minus.125\baselineskip
	\smallskipamount=.25\baselineskip
		plus.0625\baselineskip minus.0625\baselineskip
	\setbox\strutbox=\hbox{\vrule height.708\baselineskip
		depth.292\baselineskip width0pt }}



\def\makeheadline{\vbox to0pt{\baselinestretch=1000
	\vskip-\headskip \vskip1.5pt
	\line{\vbox to\ht\strutbox{}\the\headline}\vss}\nointerlineskip}

\def\makefootline{\baselineskip=\footskip\line{\the\footline}}

\def\big#1{{\hbox{$\left#1\vbox to8.5\rp@ {}\right.\n@space$}}}
\def\Big#1{{\hbox{$\left#1\vbox to11.5\rp@ {}\right.\n@space$}}}
\def\bigg#1{{\hbox{$\left#1\vbox to14.5\rp@ {}\right.\n@space$}}}
\def\Bigg#1{{\hbox{$\left#1\vbox to17.5\rp@ {}\right.\n@space$}}}


\mathchardef\alpha="710B
\mathchardef\beta="710C
\mathchardef\gamma="710D
\mathchardef\delta="710E
\mathchardef\epsilon="710F
\mathchardef\zeta="7110
\mathchardef\eta="7111
\mathchardef\theta="7112
\mathchardef\iota="7113
\mathchardef\kappa="7114
\mathchardef\lambda="7115
\mathchardef\mu="7116
\mathchardef\nu="7117
\mathchardef\xi="7118
\mathchardef\pi="7119
\mathchardef\rho="711A
\mathchardef\sigma="711B
\mathchardef\tau="711C
\mathchardef\upsilon="711D
\mathchardef\phi="711E
\mathchardef\chi="711F
\mathchardef\psi="7120
\mathchardef\omega="7121
\mathchardef\varepsilon="7122
\mathchardef\vartheta="7123
\mathchardef\varpi="7124
\mathchardef\varrho="7125
\mathchardef\varsigma="7126
\mathchardef\varphi="7127
\mathchardef\imath="717B
\mathchardef\jmath="717C
\mathchardef\ell="7160
\mathchardef\wp="717D
\mathchardef\partial="7140
\mathchardef\flat="715B
\mathchardef\natural="715C
\mathchardef\sharp="715D


\def\err@badsizechange{%
	\immediate\write16{--> Size change not allowed in math mode, ignored}}

\baselinestretch=1000
\tenpoint

\catcode`\@=12					
\catcode`\@=11
\expandafter\ifx\csname @iasmacros\endcsname\relax
	\global\let\@iasmacros=\par
\else	\immediate\write16{}
	\immediate\write16{Warning:}
	\immediate\write16{You have tried to input iasmacros more than once.}
	\immediate\write16{}
	\endinput
\fi
\catcode`\@=12


\def\rmb{\seventeenrm}

\def\singlespace{\baselineskip=\normalbaselineskip}
\def\halfspace{\baselineskip=1.5\normalbaselineskip}
\def\doublespace{\baselineskip=2\normalbaselineskip}


\def\AB{\bigskip\parindent=40pt
        \centerline{\bf ABSTRACT}\medskip\halfspace\narrower}
\def\AE{\bigskip\nonarrower\doublespace}
\def\nonarrower{\advance\leftskip by-\parindent
	\advance\rightskip by-\parindent}


\def\boxit#1{\vbox{\hrule\hbox{\vrule\kern3pt
	\vbox{\kern3pt#1\kern3pt}\kern3pt\vrule}\hrule}}

\def\hence{\leavevmode\hbox{\bf .\raise5.5pt\hbox{.}.} }

\def\dalemb#1#2{{\vbox{\hrule height.#2pt
	\hbox{\vrule width.#2pt height#1pt \kern#1pt \vrule width.#2pt}
	\hrule height.#2pt}}}
\def\gtorder{\mathrel{\raise.3ex\hbox{$>$}\mkern-14mu
             \lower0.6ex\hbox{$\sim$}}}
\def\ltorder{\mathrel{\raise.3ex\hbox{$<$}\mkern-14mu
             \lower0.6ex\hbox{$\sim$}}}

\newdimen\fullhsize
\newbox\leftcolumn
\def\twoup{\hoffset=-.5in \voffset=-.25in
  \hsize=4.75in \fullhsize=10in \vsize=6.9in
  \def\fullline{\hbox to\fullhsize}
  \let\lr=L
  \output={\if L\lr
        \global\setbox\leftcolumn=\columnbox\global\let\lr=R \advancepageno
      \else \doubleformat \global\let\lr=L\fi
    \ifnum\outputpenalty>-20000 \else\dosupereject\fi}
  \def\doubleformat{\shipout\vbox{
    \fullline{\box\leftcolumn\hfil\columnbox}\advancepageno}}
  \def\columnbox{\leftline{\vbox{\makeheadline\pagebody\makefootline}}}
  \tolerance=1000 }

\twelvepoint
\doublespace
\overfullrule=0pt
{\nopagenumbers{
\rightline{IASSNS-HEP-00/16}
\rightline{~~~March, 2000}
\bigskip\bigskip
\centerline{\rmb Scalar Exchange Forces and Generalized}
\centerline{\rmb Most Attractive Channel Rule}
\medskip
\centerline{\it Stephen L. Adler}
\centerline{\bf Institute for Advanced Study}
\centerline{\bf Princeton, NJ 08540}
\medskip
\bigskip\bigskip
\leftline{\it Send correspondence to:}
\medskip
{\singlespace\leftline{Stephen L. Adler}
\leftline{Institute for Advanced Study}
\leftline{Einstein Drive, Princeton, NJ 08540}
\leftline{Phone 609-734-8051; FAX 609-924-8399; email adler@ias.edu}}
\bigskip\bigskip
}}
\vfill\eject
\pageno=2
\AB
We discuss the possibility that fermionic condensates arise from the 
dominance of scalar exchange forces over vector gluon exchange.  When a  
scalar in the adjoint representation is exchanged in the reaction $A+A 
\to A+A$, the usual most attractive channel (MAC) rule is reversed in sign, 
with the consequence that 
the formation of a condensate with the largest possible Casimir is 
favored.  More generally, when a scalar in a general representation is 
exchanged in the reaction $A+B \to B+A$, the group theoretic sum giving 
the force sign and strength can be expressed in terms of a Racah
coefficient   
for the group in question.  We illustrate the formalism in the case 
of the group $SU(2)$, and give possible applications to $SO(10)$ and $E_6$  
grand unification.
\AE
\bigskip\bigskip
\vfill\eject
\pageno=3
Attempts to construct realistic grand unified models typically involve 
a complicated Higgs scalar sector [1],  and this has led   
to the suggestion [2] that some of these scalars may be dynamically 
generated fermionic composites. A particularly appealing scenario in this  
regard is the ``tumbling'' hypothesis [3], which suggests that a symmetry 
breaking hierarchy develops in which scalar composites, formed from the 
fermions of the theory at each level of symmetry breaking, develop vacuum  
expectation values that trigger the next stage of symmetry breaking.  
Applications of the tumbling scenario typically assume that the 
forces giving rise to condensate formation arise from vector gluon 
exchange.   However, this assumption  
leads to the problem that since the most attractive 
channel (MAC) rule [4] for vector exchange favors the 
formation of composites in 
small representations, it is not possible to generate the large Higgs 
representations needed for the construction of realistic models.  

Our purpose in this Letter is to explore the possibility that scalar 
mediated forces may play a significant role in dynamical symmetry breaking, 
and to develop an analog of the MAC rule in this case. We begin with 
a review of the MAC rule in the case in which a vector gluon mediates 
the reaction $A+B \to A +B$.  Since a gauge gluon couples with 
universal strength to the representations $A$ and $B$, the couplings 
$g_A$ and $g_B$ are equal, $g_A=g_B=g$. Thus the relevant static potential 
is 
$${g^2 \, T_A \cdot T_B \over r} ~~~,\eqno(1a)$$ 
with $\cdot$ denoting a sum over the group generators and   
with $T_{A,B}$ the generator matrices 
for the respective representations $A,B$.  Using 
$$T_A \cdot T_B = {1\over 2} [(T_A+T_B)^2-T_A^2-T_B^2]~~~,\eqno(1b)$$
together with the fact that the summed squares of the generators define 
the respective quadratic Casimir operators $C_2$, the effective force 
strength becomes 
$${g^2 [C_2(A+B)-C_2(A)-C_2(B)] \over 2r}~~~.\eqno(1c)$$   
This gives the MAC rule for vector gluon exchange, which states that the 
interaction is attractive when the square bracket in Eq.~(1c) is negative, 
and that the MAC is the channel with the smallest Casimir for the composite 
$A+B$.  

Let us consider now the case in which a scalar mediates the reaction 
$A+B \to A+B$.  Since scalar couplings are not universal and can 
have either sign, in general we have $g_A \not = g_B$, and whether the 
scalar exchange force is attractive or repulsive depends on dynamical 
details of the Yukawa couplings.  However, there are cases of physical 
interest in which a dynamics independent statement can be made.  
The simplest is that in which $A=B$, so that $g_A=g_B=g$, and in which   
the exchanged scalar is in the adjoint representation.  The calculation 
is then the same as that given above for the vector gluon case [5], except 
that the Coulomb potential $1/r$ is replaced by the 
Yukawa potential $-\exp(-\mu r) /r$, with $\mu$ the scalar mass and with 
the change in sign reflecting the fact that the Yukawa force is 
{\it attractive}, rather than repulsive, for the case of 
equal couplings at the two vertices.  Thus Eq.~(1c) becomes in this case 
$$-{g^2 [C_2(A+A)-2C_2(A)] \exp(-\mu r) \over 2r}~~~, \eqno(2)$$   
and the MAC is now the channel with the {\it largest} Casimir for the  
composite $C=A+A$.  

More generally, let us consider the reaction $A+B \to B+A$ with exchange 
of a scalar in a general representation $S$, which is emitted at a vertex 
where fermion representation $B$ changes to fermion representation $A$, 
and then is absorbed at a vertex where fermion representation $A$ changes 
to fermion representation $B$. Since the processes at the two vertices  
$B \to A +S$ and $A+S \to B$ are inverse to each other, the Yukawa coupling 
at one will be the complex conjugate of the Yukawa coupling at the other, 
and a dynamics independent statement about the sign and magnitude of the 
scalar mediated force is possible. We will be particularly interested in the 
dependence of the force on the representation of the composite $C=A+B$. 

To carry out this calculation, it is helpful to consider the general  
$S$ exchange reaction $A+B \to A^{\prime}+B^{\prime}$.  
The Bethe-Salpeter kernel corresponding to 
this process is 
$$K(A^{\prime}m_A^{\prime}, B^{\prime} m_B^{\prime}|A m_A, B m_B)~~~,
\eqno(3a)$$
with $m_{A,B},m^{\prime}_{A,B}$ the ``magnetic'' quantum numbers 
(the eigenvalues of the generators in the Cartan subalgebra) 
for the respective representations.  
The corresponding operator $K$ acting in the Hilbert space of the composites 
is then 
$$\eqalign{
K=&\sum_{m_{A,B},m^{\prime}_{A,B}}
|A^{\prime} m_A^{\prime} \rangle |B^{\prime} m_B^{\prime} \rangle
K(A^{\prime}m_A^{\prime}, B^{\prime} m_B^{\prime}|A m_A, B m_B)     
\langle A m_A| \langle B m_B|\cr
=&\sum_{C m_C,C^{\prime} m_C^{\prime}}
|C^{\prime} m_C^{\prime}\rangle K(C^{\prime} m_C^{\prime}|C m_C) 
\langle C m_C|~~~,\cr
}\eqno(3b)$$
where in the second line we have decomposed the direct product states 
into irreducible representations, adopting the convention that any indices  
that enumerate representations appearing more than once are included 
in the representation labels $C$ and $C^{\prime}$.  Since we are assuming 
that the $S$ exchange interaction is group invariant, we must have 
$$K(C^{\prime} m_C^{\prime}|C m_C)=K_C \delta_{C,C^{\prime}}
\delta_{m_C,m_C^{\prime}}~~~,\eqno(4a)$$ 
and so Eq.~(3b) takes the form 
$$K=\sum_{C,m_C} |C m_C\rangle K_C \langle C m_C |~~~.\eqno(4b)$$
Hence the eigenvalue $K_C$ which determines the force strength 
as a  function of the representation $C$ [which is 
the generalization of the 
expression $g^2[C_2(A+B)-C_2(A)-C_2(B)]$ of Eq.~(1c)] is given by 
$$K_C=\sum_{m_{A,B},m^{\prime}_{A,B}} \langle C m_C| A^{\prime}m_A^{\prime} 
B^{\prime} m_B^{\prime} \rangle 
K(A^{\prime}m_A^{\prime}, B^{\prime} m_B^{\prime}|A m_A, B m_B)     
\langle A m_A B m_B|C m_C\rangle~~~.\eqno(5a)$$

To calculate the group theoretic part of the kernel of Eq.~(3a), we note 
that there are two contributions:  one in which a vertex for the 
process $B \to B^{\prime}+S$ is joined to a vertex for $A+S \to A^{\prime}$, 
and one in which a vertex for $A \to A^{\prime}+S$ is joined to a vertex 
for $B+S \to B^{\prime}$.  Each vertex, by the Wigner-Eckart theorem, is 
the product of a Clebsch  which carries the magnetic quantum number 
dependence, times a reduced matrix element which is independent of the 
magnetic quantum numbers.  Taking account of the facts that 
the interaction Lagrangian associated with the vertex process 
$B \to B^{\prime}+S$ is the Hermitian adjoint of that associated with 
the process $B^{\prime}+S \to B$, and that the two contributions to the 
kernel are related by the interchange of the labels $A$ and $B$, we have 
$$\eqalign{ 
K(A^{\prime}m_A^{\prime}, B^{\prime} m_B^{\prime}|A m_A, B m_B)  
=&\sum_{m_S}
[g_S(B B^{\prime})^* g_S(A^{\prime} A) 
\langle B m_B|B^{\prime} m_B^{\prime} S m_S\rangle^*
\langle A^{\prime} m_A^{\prime}|A m_A S m_S\rangle\cr
+&g_S(A A^{\prime})^* g_S(B^{\prime} B) 
\langle A m_A|A^{\prime} m_A^{\prime} S m_S\rangle^*
\langle B^{\prime} m_B^{\prime}|B m_B S m_S\rangle]\cr
}~~~.\eqno(5b)$$
Self-adjointness of this kernel was assumed in the spectral analysis 
of Eqs.~(3a) - (4b), and this is manifest from the expression in 
Eq.~(5b),  
$$K^*(A m_A,B m_B|A^{\prime} m_A^{\prime}, B^{\prime} m_B^{\prime})
=K(A^{\prime} m_A^{\prime},B^{\prime} m_B^{\prime}|A m_A,B m_B)~~~.
\eqno(5c)$$
Substituting 
Eq.~(5b) into Eq.~(5a), specializing to the case of interest in which 
$A^{\prime}=B$ and $B^{\prime}=A$, and interchanging the  
summation indices 
$m_A^{\prime}$ and $m_B^{\prime}$, we get the result 
$$\eqalign{
K_C=&\sum_{m_{A,B},m^{\prime}_{A,B},m_S} \langle C m_C| B m_B^{\prime} 
A m_A^{\prime} \rangle\cr 
\times&
[|g_S(B A)|^2\langle B m_B|A m_A^{\prime} S m_S\rangle^* 
\langle B m_B^{\prime}|A m_A S m_S\rangle \cr
+&|g_S(A B)|^2\langle A m_A|B m_B^{\prime} S m_S\rangle^* 
\langle A m_A^{\prime}|B m_B S m_S\rangle]\cr
\times&\langle A m_A B m_B|C m_C\rangle ~~~.\cr
}\eqno(6a)$$
Changing to the standard notation for the Clebsches, 
$$\langle A m_A B m_B|C m_C\rangle \equiv (A m_A B m_B|A B C m_C)~~~,
\eqno(6b)$$
adopting a phase convention in which the Clebsches are real, and using 
the symmetry property 
$$(A m_A B m_B|ABC m_C)=\epsilon(A,B,C) (B m_B A m_A|BA C m_C)~~~,
\eqno(6c)$$
with $\epsilon(A,B,C)=\epsilon(B,A,C)$ a phase factor of $\pm 1$, 
Eq.~(6a) takes the form
$$
\eqalign{
K_C=&|g_S(B A)|^2 
\sum_{m_{A,B},m^{\prime}_{A,B},m_S} 
(BAC m_C| B m_B^{\prime}A m_A^{\prime} ) 
(ASB m_B^{\prime}|A m_A S m_S)\cr      
\times &
(ASB m_B|A m_A^{\prime} S m_S)^* 
(A m_A B m_B|ABC m_C)+A \leftrightarrow B\cr
=&|g_S(B A)|^2\epsilon(A,S,B) 
\sum_{m_{A,B},m^{\prime}_{A,B},m_S} 
(BAC m_C| B m_B^{\prime}A m_A^{\prime} ) 
(ASB m_B^{\prime}|A m_A S m_S)\cr      
\times &(S m_S A m_A^{\prime} |SAB m_B) 
(A m_A B m_B|ABC m_C)~+A \leftrightarrow B\cr
=&|g_S(B A)|^2\epsilon(A,S,B)((AS)B,A,C|A,(SA)B,C)\cr  
+&|g_S(A B)|^2\epsilon(B,S,A)((BS)A,B,C|B,(SB)A,C)~~~,\cr
}\eqno(7)$$
where on the final line we have introduced the standard definition [6,~7] 
of the recoupling coefficient (the Racah coefficient)  
for three group representations.  Equation (7) is our final result for   
the case of general representations $A,B,S,C$ of a general Lie group. 
When the representation $S$ is not self-conjugate, only one of the two 
reduced matrix elements on the right hand side of Eq.~(7) will be nonzero; 
when the representation $S$ is self-conjugate, there is no physical 
distinction between the two contributions to the kernel, and we expect 
symmetries of the Clebsches to collapse the two terms on the right hand 
side of Eq.~(7) into a single term.  

Let us briefly examine some special cases of this formula 
when the Lie group is $SU(2)$, for which the relevant recoupling 
coefficients are those given in the standard angular momentum texts.  
Letting $j_{A,B,C,S}$ be the angular momentum values corresponding to 
the respective representation labels $A,B,C,S$, we have 
$$\epsilon(A,S,B)=(-1)^{j_A+j_S-j_B}~,~~~\epsilon(B,S,A)=(-1)^{j_B+j_S-j_A}
~~~,\eqno(8a)$$ 
and 
$$((j_Aj_S)j_B,j_A,j_C|j_A,(j_Sj_A)j_B,j_C)=(2j_B+1)(-1)^{-(2j_A+j_S+j_C)}
\left\{\matrix{j_A&j_S&j_B\cr j_A&j_C&j_B\cr }\right\}~~~.\eqno(8b)$$   
Substituting these expressions into Eq.~(7) and using the permutation          

symmetries of the 6-j symbols, we get for $SU(2)$ 
$$K_C=[(2j_B+1)|g_S(BA)|^2+(2j_A+1)|g_S(AB)|^2]
(-1)^{-(j_A+j_B+j_C)}
\left\{\matrix{j_C&j_A&j_B\cr j_S&j_A&j_B\cr }\right\}~~~.\eqno(8c)$$   

We now use Eq.~(8c) to examine some special cases of interest.  
When the exchanged scalar has spin 0, we find from the formula given 
in Eq.~(6.3.2) of [6] that 
$$\left\{\matrix{j_C&j_A&j_B\cr 0&j_A&j_B\cr }\right\}       
=\delta_{A,B}(-1)^{j_A+j_B+j_C}
[(2j_A+1)(2j_B+1)]^{-{1\over 2}}~~~,\eqno(9a)$$
and so Eq.~(8c) vanishes when $A \not= B$, and when $A=B$ gives  
$$K_C=2|g_S(AA)|^2~~~,\eqno(9b)$$
which as one would expect is independent of the composite representation $C$.  
When the exchanged scalar has spin 1, which is the adjoint representation  
for $SU(2)$, we find from Table 5 of [6] that when $j_B=j_A$, we have 
$$\left\{\matrix{j_C&j_A&j_A\cr 1&j_A&j_A\cr }\right\}       
=(-1)^{1+2j_A+j_C}
{2j_A(j_A+1)-j_C(j_C+1) \over
j_A(2j_A+1)(2j_A+2) }~~~,\eqno(10a)$$
which gives  
$$K_C={|g_S(AA)|^2\over j_A (j_A+1)}[j_C(j_C+1)-2j_A(j_A+1)]~~~,\eqno(10b)$$ 
reproducing the $C$ dependence found from the Casimir analysis of Eq.~(2).  

The other nonvanishing case with $j_S=1$ is that with $j_B=j_A+1$; 
from Table 5 of [6] we see that in this case $K_C$ is positive and is also
monotonically increasing with the composite spin $j_C$. However, once  
we go to larger representations $S$ than the adjoint representation, it 
is {\it not} always true that $K_C$ monotonically increases with the 
size of the representation $C$.  For example, for $j_S=2$ and 
$j_A=j_B=3$, the allowed range of $j_C$ is from 0 to 6.  For these  
parameters, we have   
$$\eqalign{
&(-1)^{-(j_A+j_B+j_C)}
\left\{\matrix{j_C&j_A&j_B\cr j_S&j_A&j_B\cr }\right\} \cr 
=&(-1)^{-j_C}
\left\{\matrix{j_C&3&3\cr 2&3&3\cr }\right\} \cr
=&{360-47j_C(j_C+1)+j_C^2(j_C+1)^2 \over 2520}~~~,\cr
}\eqno(11)$$
which for the allowed range of $j_C$ assumes a maximum value of 
$1/7$ at $j_C=0$, and takes the smaller value $5/84$ at $j_C=6$.  

We turn now to a discussion of possible applications of these results 
to symmetry breaking in grand unified theories.  We begin with the 
$SO(10)$ model, with a $16$ of chiral fermions.  One of these fermions 
is a singlet under $SU(5)$, and in the ``seesaw'' mechanism of  
Gell-Mann, Ramond, and Slansky [8], is given a large mass by a 
Higgs field in the $126$ representation of $SO(10)$ that acquires a   
vacuum expectation value.  From the viewpoint of descent from 
an $E_6$ unifying group, a $126$ of $SO(10)$ is ugly, since the smallest 
$E_6$ representation [9] giving rise to it under $E_6 \to SO(10) \times U(1)$ 
is the $\overline{351^{\prime}}$.  So it is natural to consider the 
possibility that the 126 Higgs is a composite.  Let us suppose that a 
Higgs scalar in the adjoint 45 representation of $SO(10)$ is present,  
which can be obtained by descent from the adjoint 78 of $E_6$.  Since  
in $SO(10)$ we have $16 \times 45 \supset 16$, exchange of the scalar 45 
can mediate the process $16 +16 \to 16 +16$, for which the formula of 
Eq.~(2) applies. This tells us that the MAC is the channel with 
the largest Casimir appearing in $16 \times 16=10_s + 120_a + 126_s$, 
which is the $126_s$.  Since the two chiral fermion fields 
in the 16 anticommute, but have their spinor indices  
contracted with an antisymmetric two index tensor, the group theoretic  
part of the composite wave function must be symmetric, a requirement 
satisfied by the $126_s$.  We note finally that when this composite 
126 obtains an $SU(5)$ singlet vacuum expectation value, this 
can only involve the components  
of the 16 which are $SU(5)$ singlets, since none of the $SU(5)$ 
tensor products $10\times 10,~10\times \overline 5$, or $\overline 5 \times 
\overline 5$ contains an $SU(5)$ singlet.  Hence the chiral symmetries 
of the components of the $SO(10)$ 16 that are in the $\overline 5$ 
and $10$ of $SU(5)$ are preserved.  
   
We give next some possible model building applications of the more 
general analysis leading to Eq.~(7).  The first application is again to the 
generation of a composite 126 of $SO(10)$.  Since in $SO(10)$, 
$210\times 16 \supset 16$, exchange of a scalar 210 can also mediate 
the process $16+16 \to 16 +16$.  This is the $A=B$ case of Eq.~(7); 
identifying the MAC here will require computing the $SO(10)$ 
Clebsch interchange phase and Racah coefficient appearing in Eq.~(7).  
The second application concerns the possibility of generating a 45 or 210 of 
$SO(10)$, starting with an $E_6$ model containing a 27 of scalars,  
one or more chiral fermion families in the 27, and 
extra pairs $27+\overline{27}$ of chiral fermions.  Since under 
$E_6 \supset SO(10) \times U(1)$ the 27 decomposes as 
$27=1(4)+10(-2)+16(1)$, and since in $SO(10)$ we have 
$10 \times \overline{16} \supset 16$, exchange of a $10(-2)$ scalar 
can mediate the process $16(1) +\overline{16}(-1) \to 
\overline{16}(-1) + 16(1)$.  This corresponds to Eq.~(7) with 
$A=\overline{16},~B=16,~S=10$.  The possible composites $C$ are 1, 
45, and 210 of $SO(10)$, corresponding to the decomposition 
$\overline{16} \times 16 =1 +45 +210$; to identify the MAC will again 
require computation of the relevant $SO(10)$ phase factor and Racah 
coefficient. Another $SO(10) \times U(1)$ case to which our analysis  
applies is $10(-2)+10(2) \to 10(2) + 10(-2)$, which can be mediated by 
$1(4)$ exchange, and has $1_s,~45_a$, and $54_s$ as equally attractive 
composite channels.  
Our final application, which is an extension of the second,  
is to the symmetry breaking 
$E_6 \supset SO(10) \times U(1)$ in an $E_6$ model with fermion and 
scalar content as just described.  Since in $E_6$ we have $27 \times 27 
\supset \overline{27}$, exchange of a scalar 27 can mediate the process 
$27 + \overline{27} \to \overline{27} +27$, corresponding to Eq.~(7) 
with $A=27,~B=\overline{27},~S=27$.  The possible composites are 
the representations 1, 78, 650 appearing in the decomposition of $27 \times
\overline{27}$, and determining the MAC in this example will require 
computation of the $E_6$ phase factor and Racah coefficient appearing 
in Eq.~(7).  Further examples of Eq.~(7) in the same model 
can be obtained by proceeding  
down either of the symmetry breaking chains $SO(10) \to SU(5) \times U(1)$ 
or $SO(10) \to SU(2) \times SU(2) \times SU(4)$.

\bigskip
\centerline{\bf Acknowledgments}
This work was supported in part by the Department of Energy under
Grant \#DE--FG02--90ER40542.   I wish to thank J. D. Bjorken, J. Feng, 
H. D. Politzer, M. Schwartz, and F. Wilczek for informative discussions. 
\vfill\eject
\centerline{\bf References}
\bigskip
\noindent
[1]  As an example, see C. H. Albright and S. M. Barr, ``Explicit 
$SO(10)$ Supersymmetric Grand Unified Model for the Higgs and 
Yukawa Sectors'', hep-ph/0002155. \hfill\break 
\bigskip 
\noindent
[2]  L. Susskind, Phys. Rev. D20, 2619 (1979); 
S. Weinberg, Phys. Rev. D19, 1277 (1979). \hfill\break
\bigskip
\noindent
[3] S. Raby, S. Dimopoulos, and L. Susskind, Nucl. Phys. B169, 373 (1980). 
\hfill\break
\bigskip
\noindent
[4] J. M. Cornwall, Phys. Rev. D10, 500 (1974); T. Banks and S. Raby,  
Phys. Rev. D14, 2182 (1976); see also Ref. [3] above.  For an 
exposition, see M. Peskin, ``Chiral symmetry and chiral symmetry breaking'', 
in {\it Recent Advances in Field Theory and Statistical Mechanics} 
(Les Houches, 1982), J.-B. Zuber and R. Stora, eds. (North-Holland, 
Amsterdam, 1984).  \hfill\break
\bigskip
\noindent
[5]  When the fermion representations $A$ and $B$ are identical, there 
is also an exchange force term; when projected on an antisymmetrized 
composite wave function, this just doubles the contribution of the direct 
term.\hfill\break
\bigskip
\noindent
[6]  A. R. Edmonds, {\it Angular Momentum in Quantum Mechanics}, 
Princeton University Press, Princeton, 1957. \hfill\break
\bigskip
\noindent
[7]  J.-Q. Chen, P.-N. Wang, Z.-M. L\"u, and X. -B. Wu, {\it Tables of 
the Clebsch-Gordan, Racah and Subduction Coefficients of SU(n) Groups}, 
World Scientific, Singapore, 1987.\hfill\break
\bigskip
\noindent
[8]  M. Gell-Mann, P. Ramond, and R. Slansky, in {\it Supergravity}, 
P. van Nieuwenhuizen and D. Freedman, eds., North-Holland, Amsterdam, 1979, 
p. 315; T.Yanagida, in {\it Proceedings of the Workshop on the Unified 
Theory and the Baryon Number in the Universe}, O. Sawada and A. Sugamoto, 
eds., KEK Report No. 79-18, Tsukuba, 1979, p. 95; R. N. Mohapatra 
and G. Senjanovi\'c, Phys. Rev. Lett. 44, 912 (1980).  
\hfill\break
\bigskip
\noindent
[9] All quoted group theoretic results are from the tables at the end of  
R. Slansky, ``Group Theory for Model Building'', Phys. Rep. 79, 1 (1981).
\hfill\break
\bigskip
\noindent
\bye